\documentclass{aa}  

\usepackage{graphicx}
\usepackage{subfig}
\usepackage{txfonts}
\usepackage{natbib,twoopt}
\usepackage[breaklinks=true]{hyperref} %% to avoid \citeads line fills
\bibpunct{(}{)}{;}{a}{}{,}             %% natbib format for A&A and ApJ
\makeatletter
  \newcommandtwoopt{\citeads}[3][][]{\href{http://adsabs.harvard.edu/abs/#3}%
    {\def\hyper@linkstart##1##2{}%
     \let\hyper@linkend\@empty\citealp[#1][#2]{#3}}}
  \newcommandtwoopt{\citepads}[3][][]{\href{http://adsabs.harvard.edu/abs/#3}%
    {\def\hyper@linkstart##1##2{}%
     \let\hyper@linkend\@empty\citep[#1][#2]{#3}}}
  \newcommandtwoopt{\citetads}[3][][]{\href{http://adsabs.harvard.edu/abs/#3}%
    {\def\hyper@linkstart##1##2{}%
     \let\hyper@linkend\@empty\citet[#1][#2]{#3}}}
  \newcommandtwoopt{\citeyearads}[3][][]%
    {\href{http://adsabs.harvard.edu/abs/#3}
    {\def\hyper@linkstart##1##2{}%
     \let\hyper@linkend\@empty\citeyear[#1][#2]{#3}}}
\makeatother
\def\cm#1{\ifmmode {\,{\rm cm^{-#1}}}                  % cm-1, cm-2, cm-3, ...
	\else \hbox{$\,${\rm cm$^{\rm -#1}$}}\fi}
\def\raw {\ifmmode\rightarrow\else$\rightarrow$\fi}
\def\ex#1{\ifmmode {\times 10^{#1}}         % x10$^{-1}$, x10$^{-2}$, etc
	\else \hbox{{$\times 10^{\rm #1}$}}\fi}
\newcommand{\msun}{\mbox{$M_{\odot}$}}

\begin{document}

   \title{How many suns are in the sky? A SPHERE multiplicity survey of exoplanet host stars I}

   \subtitle{Four new close stellar companions including a white dwarf}

   \author{C. Ginski \inst{1,2}
          \and
          M. Mugrauer \inst{3}
          \and
          C. Adam \inst{4,5}
          \and
          N. Vogt \inst{4}
          \and
          R. G. van Holstein \inst{2,6}
          }

   \institute{Anton Pannekoek Institute for Astronomy, University of Amsterdam, Science Park 904, 1098XH Amsterdam, The Netherlands \email{c.ginski@uva.nl}
        \and Leiden Observatory, Leiden University, PO Box 9513, 2300 RA Leiden, The Netherlands
        \and Astrophysikalisches Institutund Universitäts-Sternwarte Jena, Schillergäßchen 2, D-07745 Jena, Germany
        \and
Instituto de F\'isica y Astronom\'ia, Facultad de Ciencias, Universidad de Valpara\'iso, Av. Gran Breta\~na 1111, Playa Ancha, Valpara\'iso, Chile
\and
N\'ucleo Milenio Formaci\'on Planetaria - NPF, Universidad de Valpara\'iso, Av. Gran Breta\~na 1111, Valpara\'iso, Chile
\and European Southern Observatory, Alonso de C\'{o}rdova 3107, Casilla 19001, Vitacura, Santiago, Chile \label{inst:esosantiago}
             }

   \date{Received September 15, 1996; accepted March 16, 1997}

% \abstract{}{}{}{}{} 
% 5 {} token are mandatory
 
  \abstract
  % context heading (optional)
  % {} leave it empty if necessary  
   {}
  % aims heading (mandatory)
   {We are studying the influence of stellar multiplicity on exoplanet systems, in particular systems that have been detected via radial-velocity searches. We are in particular interested in the closest companions as they would have a strong influence on the evolution of the original planet forming disks. In this study we present new companions detected during our ongoing survey of exoplanet hosts with VLT/SPHERE.}
  % methods heading (mandatory)
   {We are using the extreme adaptive optics imager SPHERE at the ESO/VLT to search for faint (sub)stellar companions. We utilized the classical coronagraphic imaging mode to perform a snapshot survey (3-6\,min integration time) of exoplanet host stars in the $K_S$-band.}
  % results heading (mandatory)
   {We detected new stellar companions to the exoplanet host stars HD\,1666, HIP\,68468, HIP\,107773, and HD\,109271. With an angular separation of only 0.38\arcsec{} (40\,au of projected separation) HIP\,107773 is among the closest companions found to exoplanet host stars. The presence of the stellar companion explains the linear radial-velocity trend seen in the system. At such a small separation the companion likely had significant influence on the evolution of the planet forming disk around the primary star.\\
   We find that the companion in the HD\,1666 system may well be responsible for the high orbit eccentricity (0.63) of the detected Jupiter class planet, making this system one of only a few where such a connection can be established. \\
   A cross-match with the Gaia DR2 catalog showed furthermore that the near infrared faint companion around HD\,109271 had been detected in the optical and is significantly brighter than in the near infrared making it a white dwarf companion. }
  % conclusions heading (optional), leave it empty if necessary 
   {}


   \maketitle
%
%-------------------------------------------------------------------

\section{Introduction}

With the discovery on an increasing number of extrasolar planets in the past decade we are in the fortunate position to have an ever increasing statistical sample, probing the outcome of the planet formation process. However, some of the properties of this sample are not yet fully characterized. One important aspect is the presence of additional stellar bodies in the system. \cite{2010ApJS..190....1R} found that close to half of all solar type stars in the Galaxy reside in binary or higher order stellar multiple systems (see also earlier results by \citealt{1976ApJS...30..273A, 1991A&A...248..485D}). Thus the influence of additional stellar companions on the planet formation process is a highly relevant question. In fact the closest known extrasolar planet to the sun, orbiting $\alpha$\,Cen\,B, is located in a stellar multiple system (\citealt{2012Natur.491..207D}).\\
In an early hydrodynamical study \cite{2000ApJ...537L..65N} found that planet formation should be inhibited in close ($\sim$50\,au), equal mass binary systems, due to the additional source of potential energy that heats up the circumstellar disk, thus preventing fragmentation of the disk. This was supported by some observational results, e.g. \cite{2011IAUS..276..409E} found that planets are less frequent in systems with additional stellar components between 35\,au and 100\,au. \\
Several observational studies investigated the influence of additional companions on the lifetimes of circumstellar disks. In particular \cite{2009ApJ...696L..84C} and \cite{2012ApJ...745...19K} found independently that the lifetime of disks in known binary systems seems to be significantly shorter (0.1-1\,Myr as opposed to the canonical $\sim$10\,Myr, e.g. \citealt{2001ApJ...553L.153H}).\\
On the other hand \cite{2008ApJ...673..477P} found with Spitzer observations, tracing the silicate emission feature at 10\,$\mu$m, that the dust evolution in young systems in Taurus is not significantly influenced by the presence of stellar companions between 10\,au and 450\,au.\\
Finally, several recent studies find that close binaries may in fact lead to an enhanced presence of giant planets (\citealt{2016ApJ...827....8N, 2019MNRAS.485.4967F}).\\
Adaptive optics (AO) imaging with large aperture telescopes is the best method to find stellar companions to exoplanet host stars at separations between a few tens of au and a few hundred au, i.e. at separations where they are not picked up by wide field surveys, but may also not be apparent in spectroscopic observations. Several such surveys have been conducted, in the past with VLT/NACO (e.g. \citealt{2007A&A...474..273E,2015MNRAS.450.3127M}) and with Keck/NIRC2 (e.g. \citealt{2015ApJ...800..138N,2015ApJ...813..130W,Ngo2017}). Recently \cite{2020A&A...635A..73B} used for the first time an extreme\footnote{For a detailed discussion of extreme adaptive optics systems we refer to \cite{2018ARA&A..56..315G}.} adaptive optics system on an 8m-class telescope to image a large sample of transiting host stars. \\
In this study we present the first results of our stellar multiplicity survey of radial-velocity exoplanet host stars using the SPHERE instrument (Spectro-Polarimetric High-contrast Exoplanet REsearch, \citealt{Beuzit2019}) at the ESO/VLT. In the course of this survey we have observed 122 systems between 2016 and 2019 with detected radial-velocity planets, making this the largest survey of its kind with an extreme adaptive optics instrument to date. The detailed results of the survey will be presented in a forthcoming publication by Vogt et al. (in prep.). Here we highlight four systems in which we have detected new stellar companions, one of which we cross-matched with the catalog of the second data release of the ESA-Gaia mission (Gaia DR2 from hereon, \citealt{2018yCat.1345....0G}) and thus identified as white dwarf. \\
In the section~\ref{sec: ind. systems} we give a brief overview of the systems where we detected these new companions, we describe the observations and the data reduction strategy as well as the astrometric and photometric extraction in sections~\ref{obs-reduc-section} and \ref{astro-phot-section}. Using the photometry of the companions and the systems age estimates we compute mass estimates in section~\ref{character-sec} and detection limits in section~\ref{detection-limits-sec}. Finally we discuss the properties of these planetary systems in the context of these new detections in section~\ref{discuss-section}.

\section{Properties of observed systems}
\label{sec: ind. systems}
% age, distance, has planet/companion, when discovered, eccentricity, semi-major axis, masses, with ref.
In the following we summarize the basic stellar parameters and previously discovered planets in our target sample. As the orbital inclinations of the planets are not known the mass estimates are always minimum masses, i.e. $m\,sin(i)$.\\
\textit{HD\,1666} is a F7 main sequence star (\citealt{1988mcts.book.....H}) located at a distance\footnote{calculated from the inverse parallax} of 118.3$\pm$0.7\,pc (Gaia DR2, \citealt{2018yCat.1345....0G}). 
\cite{2015ApJ...806....5H} reported a Jovian mass planet ($M_{\mathrm{p}}\sin i=6.4$\,M$_{\mathrm{jup}}$) around HD\,1666 with a period of $P=270$\,days and an eccentricity of $e=0.63$.
They computed the stellar parameters from isochrone fitting and found a mass of 1.50$\pm$0.07\msun\ and an age of 1.76$\pm$0.20\,Gyr. \\
\textit{HD\,109271} is an old, solar type main sequence star of spectral type G5  with a mass of 1.05$\pm$0.02\,M$_\odot$ and an age of 7.3$\pm$1.2\,Gyr (\citealt{Girardi2000}). It is located at a distance of 56.0$\pm$0.2\,pc (Gaia DR2, \citealt{2018yCat.1345....0G}).
\cite{LoCurto2013} detected two approximately Neptune mass planets (17$\pm$1\,M$_\oplus$ and 24$\pm$2\,M$_\oplus$) in the system with orbital periods of 7.9 and 30.9\,days respectively. Both of the recovered planetary signals indicate moderately eccentric orbits with an eccentricity of 0.25$\pm$0.08 for the less massive closer-in planet and 0.15$\pm$0.09 for the outer planet. The authors also speculate on the presence of a further out planet with roughly 400\,day period but can not confirm the detection with their data set.\\ 
\textit{HIP\,68468} is a solar twin, main sequence star with a spectral type of G3 (\citealt{Houk1982}). It is located at a distance of 99.9$\pm$0.7\,pc (Gaia DR2, \citealt{2018yCat.1345....0G}). \cite{Melendez2017} estimated the stellar parameters from high resolution spectra and in particular element abundances and found an age of 6.4$\pm$0.8\,Gyr. The same authors detected radial-velocity variations which were best fit by a two planet solution. One super Earth (2.9$\pm$0.8\,M$_\oplus$) on a 1.8\,day orbit and a super Neptune (26$\pm$4\,M$_\oplus$) in a much wider 194\,day orbit. Remarkably the best fit orbit for the inner super Earth is highly eccentric with an eccentricity of 0.41, while this is not the case for the outer more massive planet.  \\
\textit{HIP\,107773} is a horizontal branch giant of spectral type K1 (\citealt{Leeuwen2007}). It is located at a distance of 105.5$\pm$0.8\,pc (Gaia DR2, \citealt{2018yCat.1345....0G}). An age estimate for the star is not given in the literature, however \cite{Jones2015} find a mass of 2.42$\pm$0.27\,M$_\odot$ from a study of the available photometry of the system. Given this mass and that the star is already on the horizontal branch the age is certainly larger than 1\,Gyr. \cite{Jones2015} find radial-velocity variations that they fit with a roughly Jupiter mass planet (1.98$\pm$0.21\,M$_\mathrm{Jup}$) on a 144\,day orbit. The planet shows a small eccentricity of 0.09$\pm$0.06.
They in addition detect a significant linear trend of 14.6$\pm$1.8\,ms$^{-1}$yr$^{-1}$ in the radial-velocity that is longer than the observation period. They speculate on a second Jupiter class planet in a significantly larger orbit but could not fit a specific solution with the available data. \\
We summarize all basic parameters of the target systems and their planetary companions in table~\ref{tab:sys-properties}.

\begin{table*}
	\centering
	\caption{Summary of the properties of our target systems. We give the spectral type of the primary star, the distance and the estimated age of the system, as well as masses, periods and eccentricities of detected planetary companions.}
	\label{tab:sys-properties}
	\small
	\begin{tabular}{lccccccc}
		\hline
		System & SpTyp & d\,[pc] & age\,[Gyr] & M$_\mathrm{P} sin(i)$ & P$_\mathrm{P}$\,[d] & e$_\mathrm{P}$ \\
		\hline
		HD\,1666		& F7\,V     & 118.3$\pm$0.7     & 1.76$\pm$0.20     & 6.4$^{+0.3}_{-0.2}$\,M$_{\mathrm{Jup}}$ & 270$^{+0.8}_{-0.9}$	 & 0.63$^{+0.03}_{-0.02}$\\
		HD\,109271	    & G5\,V     & 56.0$\pm$0.2      & 7.3$\pm$1.2       & 17$\pm$1\,M$_\oplus$ & 7.8543$\pm$0.0009 & 0.25$\pm$0.08 \\
		                &      &      &        & 24$\pm$2\,M$_\oplus$ & 30.93$\pm$0.02	 & 0.15$\pm$0.09 \\
		HIP\,68468      & G3\,V     & 99.9$\pm$0.7      & 6.4$\pm$0.8       & 2.9$\pm$0.8\,M$_\oplus$ & 1.8374$\pm$0.0003	 & $\sim$0.41 \\
		                &      &      &        & 26$\pm$4\,M$_\oplus$ & 194$\pm$2	 & $\sim$0.04 \\
		HIP\,107773	    & K1\,III   & 105.5$\pm$0.8     & $>$1              & 1.98$\pm$0.21\,M$_\mathrm{Jup}$      & 144.3$\pm$0.5	 & 0.09$\pm$0.06 \\
						\hline
	\end{tabular}
\end{table*}

\section{Observations and data reduction}
\label{obs-reduc-section}

\begin{figure*}
\center
\includegraphics[width=1.0\textwidth]{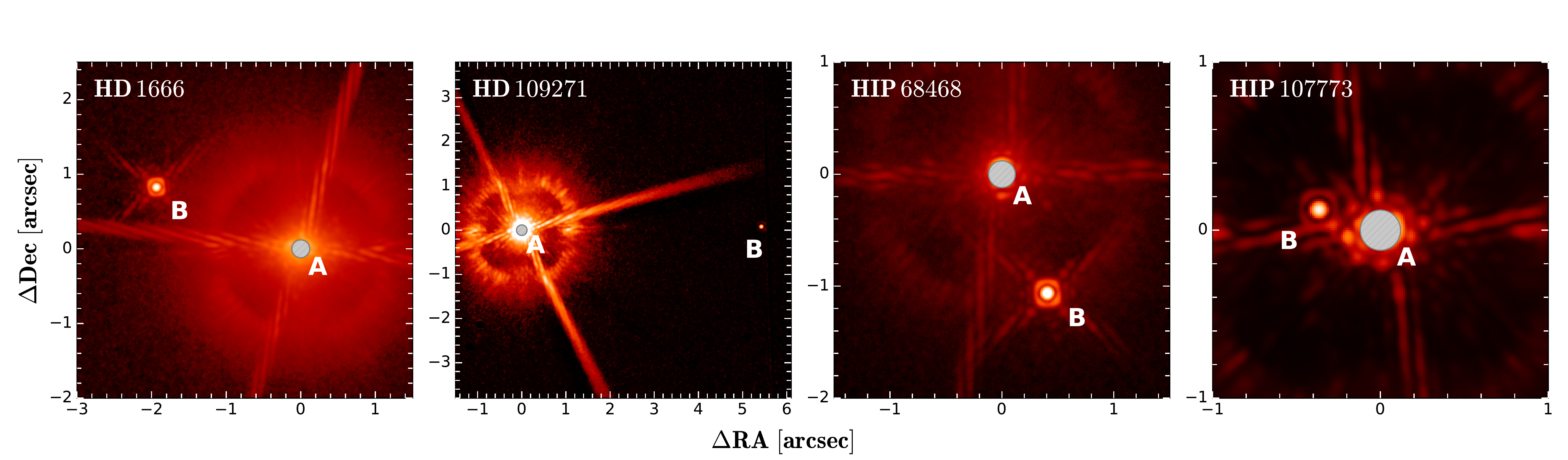} 
\caption{SPHERE/IRDIS observations of the new co-moving low-mass stellar companions to exoplanet host stars. The grey hatched disk marks the utilized coronagraphic mask. 
} 
\label{fig:sphere_images}
\end{figure*}

All observations were conducted with SPHERE/IRDIS (Infra-Red Dual Imager and Spectrograph, \citealt{Dohlen2008}) at the ESO/VLT in field stabilized classical imaging mode with the broad-band $K_s$ filter. The bright primary star was always placed behind an apodized Lyot coronagraph with an inner working angle\footnote{defined as separation with 50\,\% transmission} of 120\,mas. Individual frame exposure times (3s - 16s) were adjusted such that the residual light from the primary star does not reach saturation levels. The total integration time for all targets was between 3.2 and 6\,min. The observation conditions were highly variable between individual observation epochs and targets since our program was executed as filler for non-ideal or unstable weather conditions. We give an overview of integration times and key weather parameters in table~\ref{tab:observations}. For all systems but HD\,1666 we have at least one observation epoch with excellent seeing conditions ($<$0.8\arcsec{}) and atmosphere coherence time ($>$5\,ms). For HD\,1666 we have a high coherence time of 5.3\,ms and seeing within the tolerances of the AO system ($<$1.2\arcsec{}) in the first observation epoch in 2016 leading to a stable AO performance.\\
In all observation epochs we took flux and center reference frames as well as dedicated sky images for background subtraction. Flux calibration frames were taken with the primary star removed from the coronagraphic mask and the individual integration time adjusted to prevent saturation. Additionally neutral density filters were employed when necessary. The center reference frames were taken after the primary star was aligned behind the coronagraphic mask and the AO system was used to introduce a waffle pattern on the wavefront to create equidistant calibration spots outside of the coronagraphic mask. Sky frames were taken with the AO loop open and the telescope pointing away from the primary star with no other (bright) sources in the field of view.\\ 
For data reduction we used a modified version of the IRDAP (IRDIS Data reduction for Accurate Polarimetry, \citealt{vanHolstein2020}) pipeline. All the basic data processing steps (flat-fielding with lamp flats, sky subtraction, bad pixel masking), are executed as described in \citealt{vanHolstein2020}. Since our data was non-polarimetric we then modified the pipeline to simply center and stack all frames. The final results are shown in figure~\ref{fig:sphere_images}.

\begin{table*}
	\centering
	\caption{Observation parameters for all four systems. We give the date of observation, the integration time for a single frame (DIT) as well as the total integration time and average weather conditions (seeing and coherence time of the atmosphere).}
	\label{tab:observations}
	\small
	\begin{tabular}{lccccccc}
		\hline
		System & R.A. [hh mm ss.s] & Dec. [dd mm ss.s] & Observation Date & DIT\,[s] & Total int. time\,[s] & Seeing\,["] & Coherence time\,[ms] \\
		\hline
		HD\,1666		&	00 20 52.3 	&	-19 55 52.4 	&	25/10/2016	&	12	&	192	& 1.08		&	5.3	\\
				        &	 		    &	 		        &	05/09/2017	&	12	&	192	& 1.03		&	1.5	\\
		HD\,109271	    &	12 33 35.6 	&	-11 37 18.7 	&	29/01/2018	&	12	&	192	& 0.36		&	8.6	\\
		                &               &                   &   28/01/2019  &   12  &   192 & 1.11      &   4.5        \\
		HIP\,68468      &   14 01 03.6  &   -32 45 25.0     &   13/03/2018  &   16  &   256 & 0.90      &   7.0           \\
		                &               &                   &   20/02/2019  &   12  &   192 & 1.41      &   2.6 \\
		                &               &                   &   26/12/2019  &   16  &   192 & 0.68      &   7.0        \\
		HIP\,107773	    &	21 50 00.1 	&	-64 42 45.1 	&	09/10/2016	&	6	&	360	& 0.51		&	10.4	\\
				        &	 		    &	 		        &	04/09/2017	&	3	&	216	& 0.49		&	3.2	\\
						\hline
	\end{tabular}
\end{table*}

\section{Astrometry and photometry of new companions}
\label{astro-phot-section}

\begin{figure*}
\centering
\subfloat{
\includegraphics[width=0.9\textwidth]{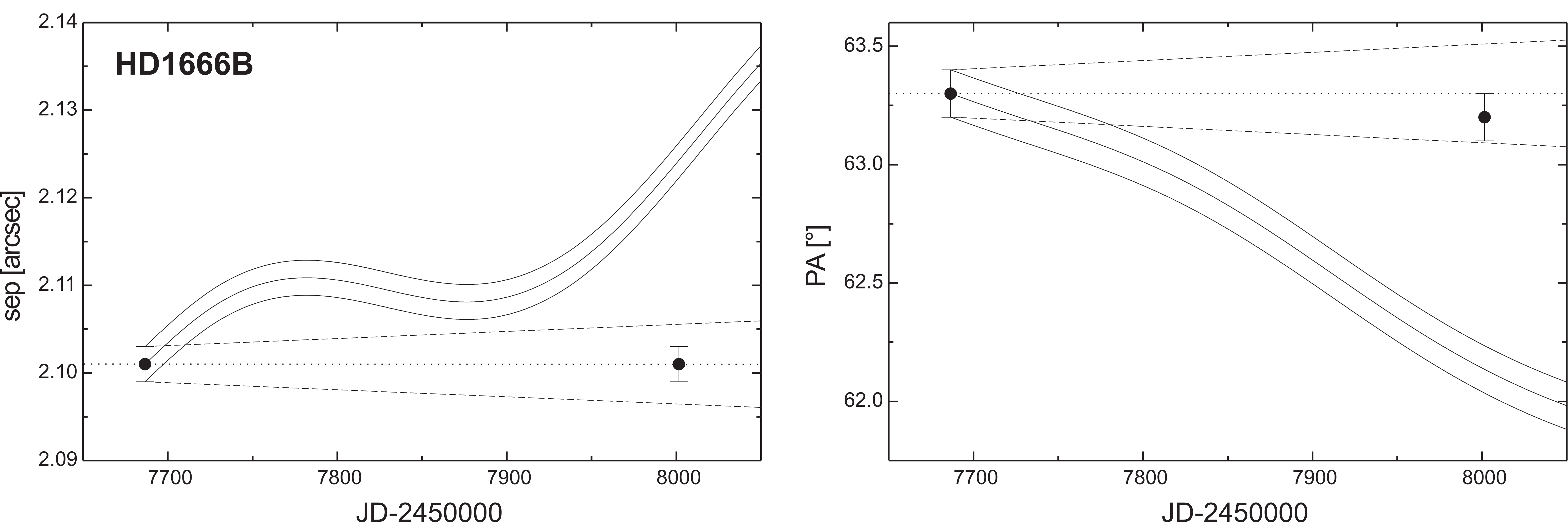}
\label{fig:astrom hd1666}
}

\subfloat{
\includegraphics[width=0.9\textwidth]{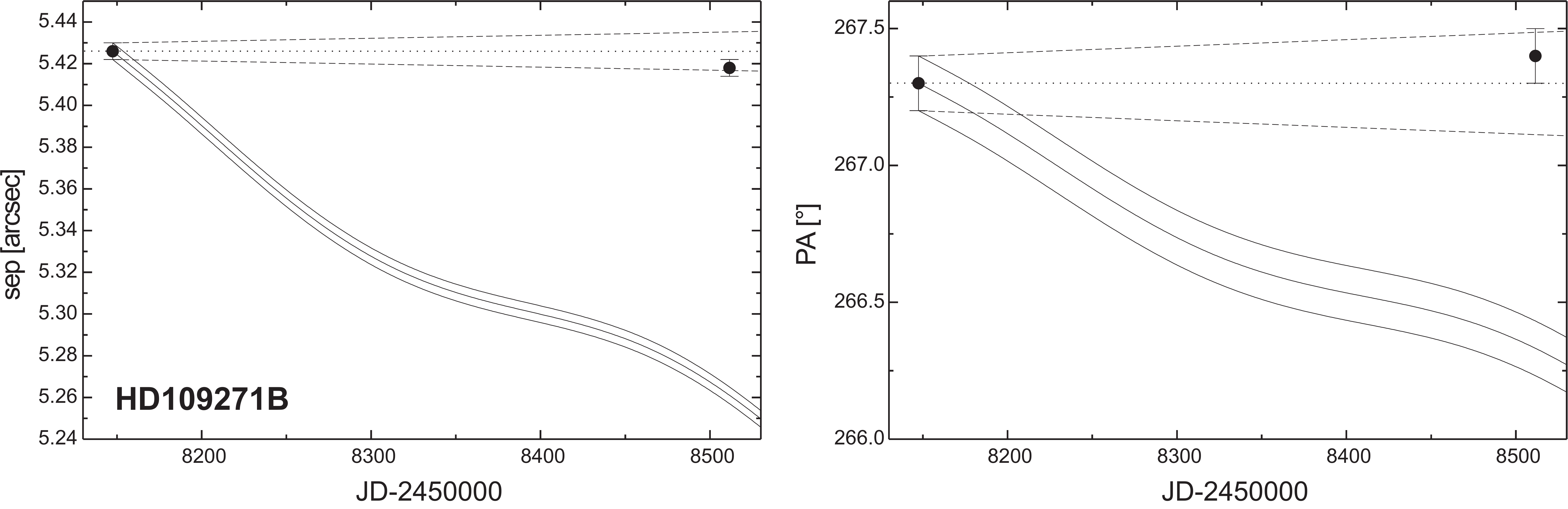}
\label{fig:astrom hd109271}
}

\subfloat{
\includegraphics[width=0.9\textwidth]{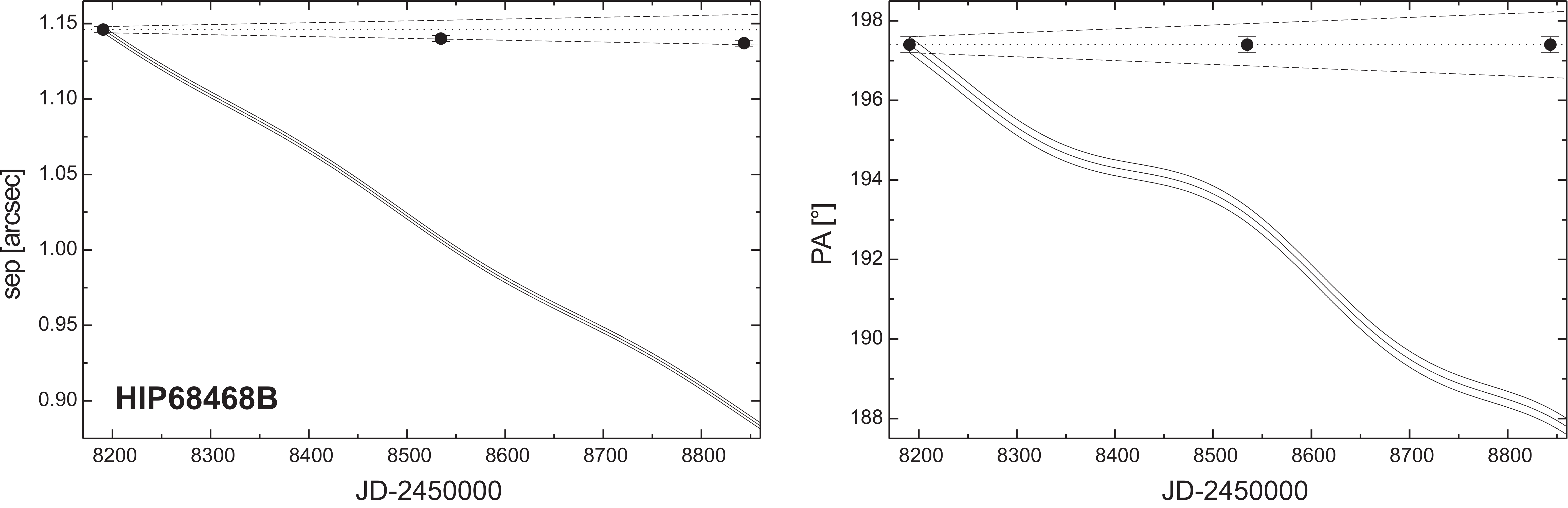}
\label{fig:astrom hip68468}
}

\subfloat{
\includegraphics[width=0.9\textwidth]{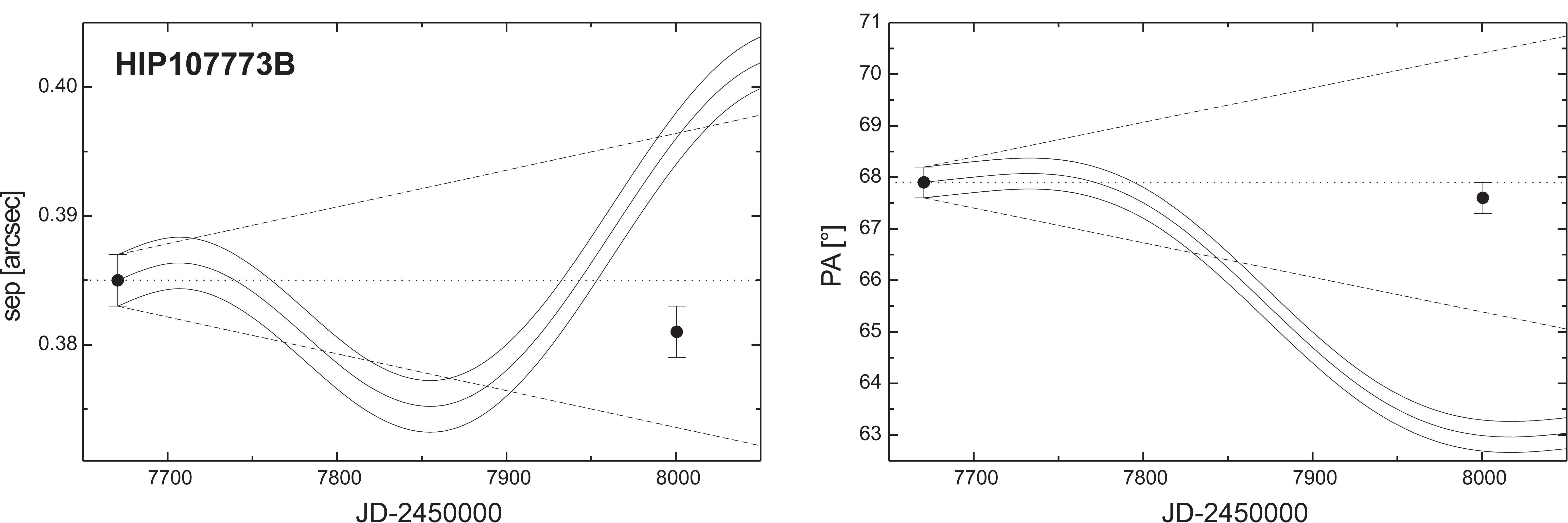}
\label{fig:astrom hip107773}
}

\caption[]{Proper motion diagrams of all detected stellar companions. The separations and position angles of the companion relative to the primary stars are plotted versus time. The "oscillating", solid lines show the area in which a non-moving background object would be expected. The dashed lines show the area where a co-moving, bound object would be expected, taking also into account possible orbital motion.}
\label{fig: astrometry}
\end{figure*}

For astrometric and photoemtric extraction we used a multi-stage fitting process which will be described in detail in van Holstein et al. (in prep.).
As a first step we fitted a Moffat function to the position of the detected companions. To account for the stellar halo of the primary star we added an inclined plane to the fit. 
We then performed PSF (point spread function) fitting, using the unsaturated PSF of the primary star from the flux calibration frames as a model. The initial guess for the peak location and scaling between primary and companion PSF was taken from the Moffat fitting results. 
The fitting procedure uses the simplex method as implemented in the \emph{minimize} function in the \emph{SciPy} package (\citealt{2020SciPy-NMeth}). To derive accurate statistical uncertainties of the fitting results we performed a Markov-Chain Monte-Carlo (MCMC) analysis using the \emph{emcee} Python package (\citealt{2013PASP..125..306F}).
We used Gaussian priors for the flux ratio between companion and primary as well as the companion position with the simplex fitting results as mean values of the Gaussians. We explored the parameter space in all cases with 32 walkers which perform a total of 20,000 steps.\\
The resulting values of the companion position in detector coordinates is converted to a relative astrometric position with respect to the primary star. For this we used the astrometric calibration given in \cite{Maire2016} for SPHERE/IRDIS, i.e. a pixel scale of 12.265$\pm$0.009\,mas/pixel and a true north detector position angle of -1.75$^\circ\pm$0.10$^\circ$. The detector position of the primary star is in all cases determined by fitting Gaussians to the satellite spots after subtraction of the residual stellar light as outlined in \cite{vanHolstein2020}. \\
The flux ratio between companion and primary star is converted to a magnitude contrast taking into account the different exposure times for flux and science frames, as well as the profiles of neutral density filters that were in some cases inserted for the flux calibration images.
The resulting astrometry and photometry is listed in table~\ref{tab:astro-phot}.
We note that the relative photometry for all companions changed significantly between epochs (changes are up to 0.9\,mag with a significance between 3$\sigma$ and 7$\sigma$). This is in all likelihood due to the unstable weather conditions of one of the observations epochs for each of the candidates. Since flux reference frames and science frames are not taken simultaneously with SPHERE, the AO corrected PSF will change between these observations. In particular during highly unstable conditions this introduces a systematic error in photometry (see e.g. \citealt{1998A&AS..129..617E} for a discussion of this effect for adaptive optics systems).
As we report in table~\ref{tab:observations}, for HD\,1666 the second epoch coherence time was well below the limit of 3\,ms after which the SPHERE AO system can not keep up with atmosphere changes well anymore. For HD\,109271 the seeing in the second observation epoch was close to the specification limit of SPHERE at 1.2\arcsec{}. Additionally thin clouds were reported in both observation epochs of HD\,109271, which may have introduced a variable sky transparency.
For HIP\,68468 the photometry of the first and third observation epoch are in perfect agreement. Both of these epochs had above average atmosphere coherence times of 7\,ms\footnote{Coherence times longer than 5.2\,ms are achieved in less than 10\,\% of the available observation time, see: \href{https://www.eso.org/sci/observing/phase2/ObsConditions.SPHERE.html}{https://www.eso.org/sci/observing/phase2/ObsConditions.SPHERE.html}} and good seeing conditions.
The second observation epoch of varies significantly from the other two, but was taken under average seeing of 1.4\arcsec{}, i.e. well outside of the SPHERE specification limit and with coherence times shorter than 3\,ms. 
Finally HIP\,107773 shows a 3.1$\sigma$ variation between both observation epochs, despite good average seeing conditions and atmosphere coherence times above 3\,ms reported in table~\ref{tab:observations}. However, in the second observation epoch the seeing and coherence time degraded significantly during the between science and flux calibration exposures. While the average seeing during the science sequence was 0.49\arcsec{}, it degraded to 0.79\arcsec{} during the flux calibration. The atmosphere coherence time dropped at the same time to 2.3\,ms, i.e. below the threshold at which the SPHERE AO has problems keeping up with the changes in atmospheric distortions.\\  
 To test if the different magnitudes may indeed be explained by a degradation in AO performance we performed aperture photometry in the second epoch of the HD\,1666 system, using a large aperture radius of 30 pixels to include all flux of the companion. Using this technique we arrive at a brighter companion contrast of 5.1$\pm$0.1\,mag compared to the 5.70$\pm$0.31\, mag extracted with PSF fitting photometry for the same data set. This result is in better agreement with the magnitude difference of 4.83$\pm$0.07\,mag measured in the first epoch under significantly better weather conditions.
The remaining discrepancy to the first epoch can be explained if some signal of the companion dropped below the noise floor in the image due to the lower quality AO correction. 
Since for all systems the first observation epoch was taken in excellent atmospheric conditions we adopt these magnitudes for the subsequent analysis, but report all results in table~\ref{tab:astro-phot} for completeness.\\
To confirm that the newly detected companions are bound to the observed host stars we performed a common proper motion analysis. Proper motions for all systems were taken from the Gaia DR2 catalog.
In Fig.~\ref{fig: astrometry} we show for each system separation and position angle of the companion relative to the primary star versus time. The solid, oscillating lines indicate the expected position for a distant and thus non-moving background object given the primary stars proper and parallactic motion and the position of the companion in the first observing epoch. We find that in all four systems the astrometry is inconsistent with such a background object with high significance. In all systems the extracted astrometry is consistent with primary star and companion exhibiting the same proper motion on the sky. Given our small field of view and that all stars in our study are evolved and not part of young co-moving groups we thus conclude that we identified in all cases new and gravitationally bound companions to these exoplanet host stars.  

\begingroup
\setlength{\tabcolsep}{5.0pt} % Default value: 6pt
\renewcommand{\arraystretch}{1.0} % Default value: 1
\begin{table}
	\centering
	\caption{Astrometric and photometric measurements (in $K_s$-band) of companions}
	\label{tab:astro-phot}
	\small
	\begin{tabular}{lcccc}
		\hline
		System 	& Epoch [yr] & Sep ["] & PA [$^\circ$] & $\Delta$mag \\
		\hline
		%HD1666    & 2016.817 & 2.099$\pm$0.004 & 63.3$\pm$0.1  & 4.83$\pm$0.07 \\
		%	        & 2017.680 & 2.099$\pm$0.004 & 63.2$\pm$0.1  & 5.70$\pm$0.31 \\
		
		HD1666    & 2016.817 & 2.101$\pm$0.002 & 63.3$\pm$0.1  & 4.83$\pm$0.07 \\
			        & 2017.680 & 2.101$\pm$0.002 & 63.2$\pm$0.1  & 5.70$\pm$0.31 \\
			        
		%HD109271  & 2018.079 & 5.427$\pm$0.009 & 267.3$\pm$0.1 & 9.57$\pm$0.03 \\
		%            & 2019.077 & 5.417$\pm$0.009 & 267.4$\pm$0.1 & 10.45$\pm$0.13 \\
		            
		HD109271  & 2018.079 & 5.426$\pm$0.004 & 267.3$\pm$0.1 & 9.57$\pm$0.03 \\
		            & 2019.077 & 5.418$\pm$0.004 & 267.4$\pm$0.1 & 10.45$\pm$0.13 \\            
		%HIP68468  & 2018.199 & 1.147$\pm$0.002 & 197.5$\pm$0.1 & 3.83$\pm$0.03 \\
		%            & 2019.139 & 1.140$\pm$0.002 & 197.5$\pm$0.1 & 3.73$\pm$0.01  \\
		%            & 2019.986 & 1.137$\pm$0.002 & 197.5$\pm$0.1 & 3.83$\pm$0.01 \\
		            
		HIP68468  & 2018.199 & 1.146$\pm$0.002 & 197.4$\pm$0.2 & 3.83$\pm$0.03 \\
		            & 2019.139 & 1.140$\pm$0.002 & 197.4$\pm$0.2 & 3.73$\pm$0.01  \\
		            & 2019.986 & 1.137$\pm$0.002 & 197.4$\pm$0.2 & 3.83$\pm$0.01 \\            
		%HIP107773 & 2016.773 & 0.383$\pm$0.001 & 68.0$\pm$0.1  & 6.75$\pm$0.01 \\
		%	        & 2017.678 & 0.377$\pm$0.001 & 68.2$\pm$0.1  & 6.97$\pm$0.07 \\
		
		HIP107773 & 2016.773 & 0.385$\pm$0.002 & 67.9$\pm$0.3  & 6.75$\pm$0.01 \\
			        & 2017.678 & 0.381$\pm$0.002 & 67.6$\pm$0.3  & 6.97$\pm$0.07 \\
		
		\hline
	\end{tabular}
\end{table}
\endgroup

\section{Characterization of companions}
\label{character-sec}

\begin{figure}[h!]
\resizebox{\hsize}{!}{\includegraphics{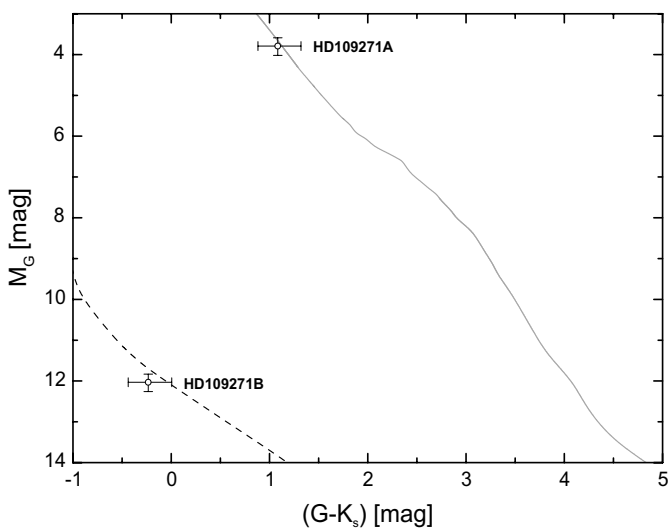}}\caption{Both components of the HD\,109271 system plotted in a $(G-K_s)-M_G$ diagram. The main-sequence is shown as grey line, the evolutionary track of DA white dwarfs with a mass of 0.6\,$M_\odot$ as black dashed line, respectively. The photometry of the exoplanet host star HD\,109271\,A is consistent with a main-sequence star, as expected. In contrast the detected co-moving companion HD\,109271\,B is clearly located below the main-sequence but its photometry agrees well with that of DA white dwarfs.}\label{FIG1}
\end{figure}

We used the extracted photometry together with the system ages and distances from section~\ref{sec: ind. systems}, as well as the 2MASS (\citealt{2006AJ....131.1163S}) $K$-band magnitudes of the primary stars to compute mass estimates for the companions.
We first converted the magnitude differences from table~\ref{tab:astro-phot} into apparent $K$-band magnitudes using the primary star $K$-band magnitude. In all cases we use the epoch with the smaller uncertainties (and better weather conditions) for the mass estimates. Given the apparent magnitudes we convert these to absolute magnitudes using the known distances and then compare these with (sub)stellar BT-SETTL model isochrones (\citealt{2015A&A...577A..42B}). To derive masses we interpolate the model grid and take into account uncertainties in all parameters, i.e. age, distance, magnitude.
We find companion masses of 0.39$\pm$0.01\,M$_\odot$ for HD\,1666\,B, 0.36$\pm$0.01\,M$_\odot$ for HIP\,68468\,B and 0.63$\pm$0.04\,M$_\odot$ for HIP\,107773\,B. We summarize the masses and projected separations of all companions in table~\ref{tab:mass-sep}.\\
The new companion in the HD\,109271 system is in principle faint enough in the $K_s$-band to be in the brown dwarf mass range. Since it is located at a very wide separation of 5.4\arcsec{} we cross-checked the Gaia DR2 catalog to see if it was picked up by Gaia as well. We indeed find that the object with Gaia identifier Gaia\,DR2\,3578137911427752704 is located at the correct separation and position angle relative to HD\,109271 (separation of 5.4250\arcsec{}$\pm$0.0007\arcsec{} and position angle of 267.354$^\circ\pm$0.004$^\circ$). This object shows a Gaia magnitude of 16.125$\pm$0.009\,mag. Given the very faint $K_s$-band magnitude of 16.06$\pm$0.04\,mag this magnitude in the optical Gaia band is surprising. If we only take our $K_s$-band measurement we find that the object should be a brown dwarf with a mass of 72.9$\pm$0.3\,M$_\mathrm{Jup}$. In this case we would expect a G-band magnitude of roughly 21.8\,mag. The significantly brighter G-band magnitude points however to a different nature of the object. \\
We use the G-band photometry of both components of the HD\,109271 system, as well as the parallax ($\pi=17.8697\pm0.066$\,mas) and the G-band extinction estimate ($A_G=0.3537_{-0.2328}^{+0.1973}$\,mag) of the primary, listed in the Gaia DR2, to determine the absolute G-band magnitudes of both stars. We obtain $M_G=3.79_{-0.20}^{+0.23}$\,mag for HD\,109271\,A, and $M_G=12.03_{-0.20}^{+0.23}$\,mag for its co-moving companion, respectively.\\
Furthermore, with the G-band extinction of the primary its 2MASS $K_s$-band magnitude ($K_s=6.495\pm0.026$\,mag) as well as the photometry of the companion, as measured in our SPHERE images, we derive the intrinsic $(G-K_s)$ color of both stars by adopting $A_{K_s} = (0.12/0.77) A_G$. This yields $(G-K_s)_{0}=1.09_{-0.21}^{+0.24}$\,mag for the primary and $(G-K_s)_{0}=-0.23_{-0.21}^{+0.24}$\,mag for the fainter secondary, respectively.\\
The derived photometry of both components of the HD\,109271 system is illustrated in a color-magnitude diagram in Fig.\,\ref{FIG1}. The stars are plotted in this diagram together with the main-sequence (grey line) from \cite{pecaut2013}\footnote{Online available in its latest version at:  \href{http://www.pas.rochester.edu/~emamajek/EEM_dwarf_UBVIJHK_colors_Teff.txt}{\url{http://www.pas.rochester.edu/~emamajek/EEM_dwarf_UBVIJHK_colors_Teff.txt}}}, as well as the evolutionary track (dashed black line) of 0.6\,$M_\odot$ DA white dwarfs, as predicted by the models of \cite{holberg2006}, \cite{kowalski2006}, \cite{tremblay2011}, and \cite{bergeron2011}.
While the photometry of HD\,109271\,A is fully consistent with that expected for a main-sequence star the companion is clearly located below the main-sequence but its photometry agrees well with that of DA white dwarfs. Hence, we conclude that HD\,109271\,B is a white dwarf companion of the exoplanet host star. Spectroscopic follow-up observations are needed to further constrain the properties of this degenerated star.\\
\cite{Mugrauer2019} computed detection limits for low-mass companions around a solar-mass, main-sequence star within a distance of 240\,pc. They find that around these targets Gaia is generally not sensitive to companions inside an angular separation of 1\arcsec{} and has detection limits corresponding to masses of 0.5$\,$M$_\odot$ inside of 2\arcsec{}.
Accordingly the other three detected companions are at too small separation and are too low-mass (and thus too faint) to be detected by Gaia.  

\begin{table}
	\centering
	\caption{Mass estimates and projected separations}
	\label{tab:mass-sep}
	\small
	\begin{tabular}{lcc}
		\hline
		Object 	& Mass [M$_\odot$] & Proj. sep [au] \\
		\hline
		HD\,1666\,B   & 0.39$\pm$0.01\,$M_\odot$ & 248 \\
		HD\,109271\,B & $\sim$0.6\,$M_\odot$ & 304 \\
		HIP\,68468\,B & 0.36$\pm$0.01\,$M_\odot$ & 114 \\
		HIP\,107773\,B & 0.63$\pm$0.04\,$M_\odot$ & 40\\
		\hline
	\end{tabular}
\end{table}

\section{Detection limits}
\label{detection-limits-sec}

\begin{figure}
\center
\includegraphics[width=0.48\textwidth]{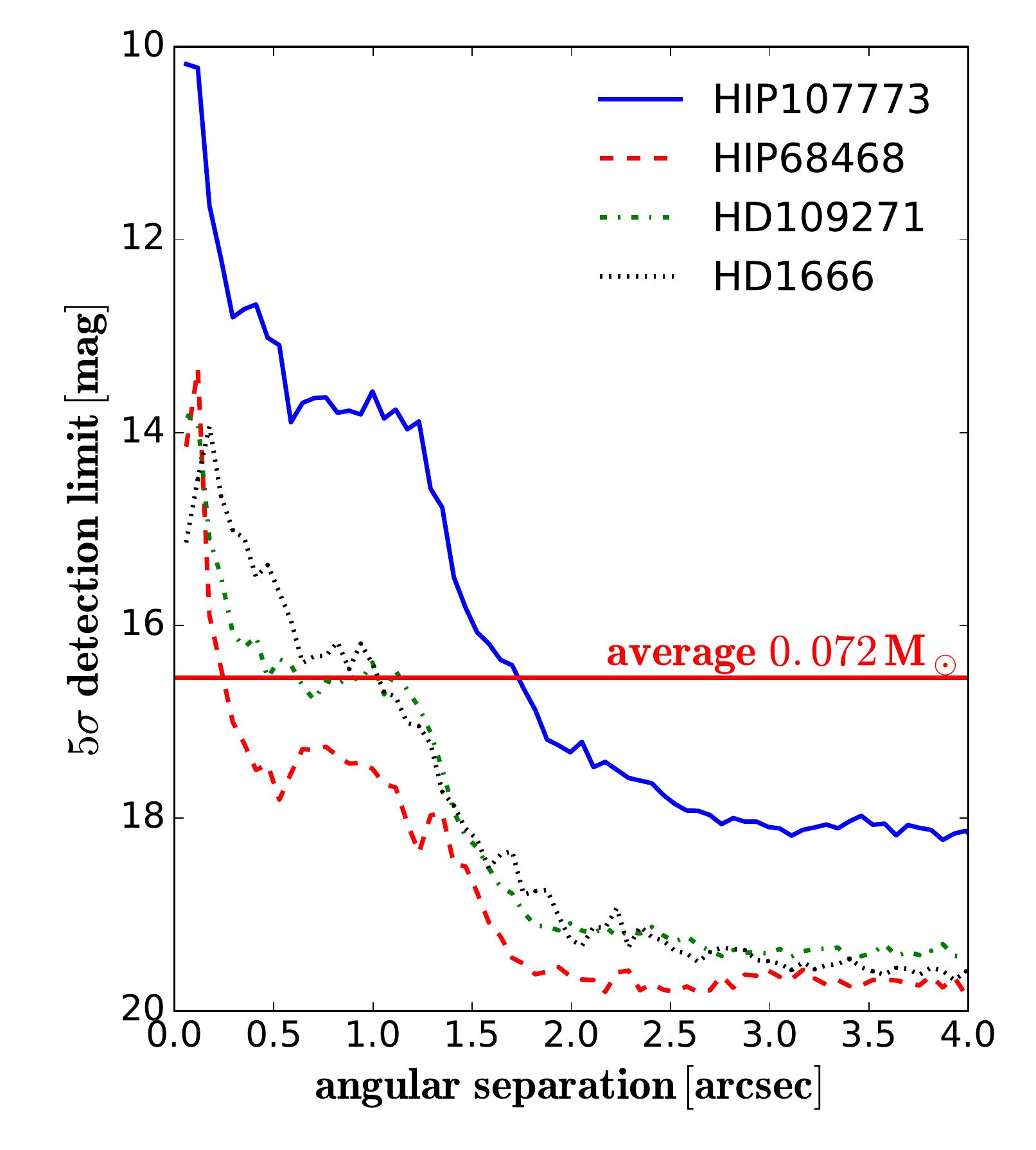} 
\caption{5\,$\sigma$ detection limits for possible companions around the four discussed systems of our survey. We show the average magnitude limit (assuming a distance of 95\,pc and and age of 4.5\,Gr), which translates into a companion mass at the stellar-brown dwarf boarder.
} 
\label{fig:contrast}
\end{figure}

While a detailed analysis of the detection limits for our entire survey sample will be presented in Vogt et al. (in prep.), we did compute individual detection limits for the four systems presented here. 
We determined the attained contrast as a function of separation using star-centered rings of comparisons, similar to the method described in \cite{2014ApJ...792...97M}, and implemented in IRDAP. Using the 2MASS magnitudes of the systems the contrast as then converted to the apparent magnitude detection limit.
The result is shown in figure~\ref{fig:contrast}.
Using the system ages and distances we can convert these magnitude limits to mass limits. For the conversion we used (sub)stellar model isochrones by \cite{2015A&A...577A..42B}. 
For HD\,109271 and HIP\,68468 we can rule out additional stellar companions down to 0.15\arcsec{}. For HD\,1666 this separation is increased to 0.4\arcsec{}. For HIP\,107773, the oldest system in this sample, we can rule out stellar companion down to 1.3\arcsec{}. At larger separations, outside of 3\arcsec{}, we are on average sensitive to wide brown dwarf companions with masses down to $\sim$60$M_\mathrm{Jup}$. Due to the high system ages we are not sensitive to objects in the planetary mass range, i.e. below the deuterium burning mass limit. 

\section{Discussion and conclusion}
\label{discuss-section}

In this study we present the discovery of four new stellar companions to exoplanet host stars one of which was identified as a white dwarf.
Of the discovered stellar companions HIP\,107773\,B is certainly the most interesting object located at a projected separation of only 40\,au.
HIP\,107773\,B is adding to a small sample of only $\sim$10\,systems with known extrasolar planets and stellar companions at orbital separations smaller than 100\,au (see \citealt{2015pes..book..309T} for a summary of these systems and \cite{Ngo2017} for updated statistics).  At such small separations the stellar companion had likely significant influence on the evolution of the planet forming disk around the primary star. In particular \cite{2018MNRAS.473.5630R} report that for companion separations of 20-30\,au the disk dispersal mechanism around the primary component changes from the photo-evaporation dominated (inside-out) regime to the tidal torque driven (outside-in) regime. 
\cite{2012A&A...542A..92R} found that systems with close stellar companions tend to harbor more massive planets, i.e. they find that for stellar separations of $\sim$40\,au all planetary masses were above 0.3\,M$_{Jup}$. Indeed the minimum mass of the detected RV planet in the HIP\,107773 system is $\sim$2\,M$_{Jup}$, which fits into this picture.
Besides the Jupiter-mass planet in the system, \cite{Jones2015} find a linear trend in the RV data. We investigated if this signal may be caused by the detected stellar companion. At its current projected separation of 40\,au the orbital period would be of the order of 145\,years. Assuming a circular orbit and an inclination of 90$^\circ$ of the orbital plane we find a semi-amplitude of the induce RV signal in HIP\,107773\,A of 2.1\,km/s. The linear trend observed by \cite{Jones2015} is changing the radial-velocity of the A component by several 10\,m/s. Thus in principle the stellar companion can well explain this signal, especially if it is currently close to the projected apastron and thus the induced signal is small, or if the inclination of the orbit is much smaller than 90$^\circ$. 
For the other three systems in this study the projected separations are much larger and thus the expected RV signal is much smaller and consistent with not being detected in the existing radial-velocity data. \\
The white dwarf companion around HD\,109271 is one of a growing number of such objects detected around exoplanet hosts (see e.g. \cite{Mugrauer2019}, who reports 9 new white dwarf companions). It appears increasingly common that planets around the less massive component in a binary system survive the post main sequence phase of the more massive component.\\
For the HD\,1666 system there is some indication that the stellar companion may have an influence on the orbital dynamics of the detected long period, super Jupiter in the system, which shows a very high eccentricity of 0.63. It may in principle be possible that HD\,1666\,b and B are locked in a Kozai-Lidov type resonance (\citealt{1962AJ.....67..591K}). Using the formula by \cite{2000ApJ...535..385F} we estimate that the period for such a resonance would be on the order of 43\,Myr if the stellar companion is currently on a circular orbit or it may be as low as 3.5\,Myr if the companion orbit is highly eccentric (e=0.9). Given that the system age is above 1\,Gyr such resonances are thus in principle possible and indeed the HD\,1666 system occupies a parameter space for which \cite{2005ApJ...627.1001T} found that Kozai-Lidov type oscillations should be effective. However, we note that \cite{Ngo2017} did not find a systematic difference between orbital parameters of planets located between 0.1\,au and 5\,au from the primary star in stellar multiple and single star systems. This may make HD\,1666 an exceptional case.\\
The parameter space probed by our new study using VLT/SPHERE in classical imaging mode is significantly different than other studies and complements those. \cite{Mugrauer2019} has recently demonstrated that, using Gaia DR2, wide stellar companions to exoplanet host stars, which are of similar brightness as the primary star, can be discovered in principle down to 1\arcsec{} (few tens of au). However, the majority of the systems that were picked up with Gaia are at separations larger than 1000\,au. With SPHERE we probe significantly closer to the primary star and for fainter and thus lower mass objects and cover a complementary parameter space compared to Gaia. The three main-sequence stellar companions that we recovered are located on projected separations between 40\,au and 250\,au and are not listed in the Gaia DR2 catalog.
In principle our imaging data is sensitive to all stellar companions down to the hydrogen burning mass limit for minimum projected separations ranging between 8\,au and 140\,au, depending on the system ages and distances. SPHERE with its extreme adaptive optics system enables thus to probe closer to the exoplanet host stars than was done previously with adaptive optics surveys, e.g. the contrast limit reached at 1.5\arcsec is roughly 2-4\,mag deeper than the observations reported in \cite{Ngo2017} with Keck/NIRC2 in the K-band. This highlights the necessity for large surveys of exoplanet host stars with extreme adaptive optics instruments, to extend our picture of stellar multiplicity in exoplanet host stars to the smallest separations and companion masses.

\begin{acknowledgements}
      SPHERE is an instrument designed and built by a consortium
consisting of IPAG (Grenoble, France), MPIA (Heidelberg, Germany),
LAM (Marseille, France), LESIA (Paris, France), Laboratoire Lagrange
(Nice, France), INAF - Osservatorio di Padova (Italy), Observatoire de
Genève (Switzerland), ETH Zurich (Switzerland), NOVA (Netherlands), ONERA
(France), and ASTRON (The Netherlands) in collaboration with ESO.
SPHERE was funded by ESO, with additional contributions from CNRS
(France), MPIA (Germany), INAF (Italy), FINES (Switzerland), and NOVA
(The Netherlands). SPHERE also received funding from the European Commission
Sixth and Seventh Framework Programmes as part of the Optical Infrared
Coordination Network for Astronomy (OPTICON) under grant number RII3-Ct2004-001566
for FP6 (2004-2008), grant number 226604 for FP7 (2009-2012),
and grant number 312430 for FP7 (2013-2016). 
C.G. acknowledges funding from the Netherlands Organisation for Scientific Research (NWO) TOP-1 grant as part
of the research program “Herbig Ae/Be stars, Rosetta stones for understanding
the formation of planetary systems”, project number 614.001.552.
This research has used the SIMBAD database, operated at CDS, Strasbourg, France \citep{Wenger2000}. 
We used the \emph{Python} programming language\footnote{Python Software Foundation, \url{https://www.python.org/}}, especially the \emph{SciPy} \citep{2020SciPy-NMeth}, \emph{NumPy} \citep{oliphant2006guide}, \emph{Matplotlib} \citep{Matplotlib}, \emph{photutils} \citep{photutils}, and \emph{astropy} \citep{astropy_1,astropy_2} packages.
We thank the writers of these software packages for making their work available to the astronomical community.
\end{acknowledgements}

% WARNING
%-------------------------------------------------------------------
% Please note that we have included the references to the file aa.dem in
% order to compile it, but we ask you to:
%
% - use BibTeX with the regular commands:
%   \bibliographystyle{aa} % style aa.bst
%   \bibliography{Yourfile} % your references Yourfile.bib
%
% - join the .bib files when you upload your source files
%-------------------------------------------------------------------

\bibliographystyle{aa}
\bibliography{MyBibFMe}

\begin{thebibliography}{56}
\expandafter\ifx\csname natexlab\endcsname\relax\def\natexlab#1{#1}\fi

\bibitem[{{Abt} \& {Levy}(1976)}]{1976ApJS...30..273A}
{Abt}, H.~A. \& {Levy}, S.~G. 1976, \apjs, 30, 273

\bibitem[{{Astropy Collaboration} {et~al.}(2018){Astropy Collaboration},
  {Price-Whelan}, {Sip{\H{o}}cz}, {G{\"u}nther}, {Lim}, {Crawford}, {Conseil},
  {Shupe}, {Craig}, {Dencheva}, {Ginsburg}, {Vand erPlas}, {Bradley},
  {P{\'e}rez-Su{\'a}rez}, {de Val-Borro}, {Aldcroft}, {Cruz}, {Robitaille},
  {Tollerud}, {Ardelean}, {Babej}, {Bach}, {Bachetti}, {Bakanov}, {Bamford},
  {Barentsen}, {Barmby}, {Baumbach}, {Berry}, {Biscani}, {Boquien}, {Bostroem},
  {Bouma}, {Brammer}, {Bray}, {Breytenbach}, {Buddelmeijer}, {Burke},
  {Calderone}, {Cano Rodr{\'\i}guez}, {Cara}, {Cardoso}, {Cheedella}, {Copin},
  {Corrales}, {Crichton}, {D'Avella}, {Deil}, {Depagne}, {Dietrich}, {Donath},
  {Droettboom}, {Earl}, {Erben}, {Fabbro}, {Ferreira}, {Finethy}, {Fox},
  {Garrison}, {Gibbons}, {Goldstein}, {Gommers}, {Greco}, {Greenfield},
  {Groener}, {Grollier}, {Hagen}, {Hirst}, {Homeier}, {Horton}, {Hosseinzadeh},
  {Hu}, {Hunkeler}, {Ivezi{\'c}}, {Jain}, {Jenness}, {Kanarek}, {Kendrew},
  {Kern}, {Kerzendorf}, {Khvalko}, {King}, {Kirkby}, {Kulkarni}, {Kumar},
  {Lee}, {Lenz}, {Littlefair}, {Ma}, {Macleod}, {Mastropietro}, {McCully},
  {Montagnac}, {Morris}, {Mueller}, {Mumford}, {Muna}, {Murphy}, {Nelson},
  {Nguyen}, {Ninan}, {N{\"o}the}, {Ogaz}, {Oh}, {Parejko}, {Parley}, {Pascual},
  {Patil}, {Patil}, {Plunkett}, {Prochaska}, {Rastogi}, {Reddy Janga},
  {Sabater}, {Sakurikar}, {Seifert}, {Sherbert}, {Sherwood-Taylor}, {Shih},
  {Sick}, {Silbiger}, {Singanamalla}, {Singer}, {Sladen}, {Sooley},
  {Sornarajah}, {Streicher}, {Teuben}, {Thomas}, {Tremblay}, {Turner},
  {Terr{\'o}n}, {van Kerkwijk}, {de la Vega}, {Watkins}, {Weaver}, {Whitmore},
  {Woillez}, {Zabalza}, \& {Astropy Contributors}}]{astropy_2}
{Astropy Collaboration}, {Price-Whelan}, A.~M., {Sip{\H{o}}cz}, B.~M., {et~al.}
  2018, \aj, 156, 123

\bibitem[{{Astropy Collaboration} {et~al.}(2013){Astropy Collaboration},
  {Robitaille}, {Tollerud}, {Greenfield}, {Droettboom}, {Bray}, {Aldcroft},
  {Davis}, {Ginsburg}, {Price-Whelan}, {Kerzendorf}, {Conley}, {Crighton},
  {Barbary}, {Muna}, {Ferguson}, {Grollier}, {Parikh}, {Nair}, {Unther},
  {Deil}, {Woillez}, {Conseil}, {Kramer}, {Turner}, {Singer}, {Fox}, {Weaver},
  {Zabalza}, {Edwards}, {Azalee Bostroem}, {Burke}, {Casey}, {Crawford},
  {Dencheva}, {Ely}, {Jenness}, {Labrie}, {Lim}, {Pierfederici}, {Pontzen},
  {Ptak}, {Refsdal}, {Servillat}, \& {Streicher}}]{astropy_1}
{Astropy Collaboration}, {Robitaille}, T.~P., {Tollerud}, E.~J., {et~al.} 2013,
  \aap, 558, A33

\bibitem[{{Baraffe} {et~al.}(2015){Baraffe}, {Homeier}, {Allard}, \&
  {Chabrier}}]{2015A&A...577A..42B}
{Baraffe}, I., {Homeier}, D., {Allard}, F., \& {Chabrier}, G. 2015, \aap, 577,
  A42

\bibitem[{{Bergeron} {et~al.}(2011){Bergeron}, {Wesemael}, {Dufour},
  {Beauchamp}, {Hunter}, {Saffer}, {Gianninas}, {Ruiz}, {Limoges}, {Dufour},
  {Fontaine}, \& {Liebert}}]{bergeron2011}
{Bergeron}, P., {Wesemael}, F., {Dufour}, P., {et~al.} 2011, ApJ, 737, 28

\bibitem[{{Beuzit} {et~al.}(2019){Beuzit}, {Vigan}, {Mouillet}, {Dohlen},
  {Gratton}, {Boccaletti}, {Sauvage}, {Schmid}, {Langlois}, {Petit},
  {Baruffolo}, {Feldt}, {Milli}, {Wahhaj}, {Abe}, {Anselmi}, {Antichi},
  {Barette}, {Baudrand}, {Baudoz}, {Bazzon}, {Bernardi}, {Blanchard}, {Brast},
  {Bruno}, {Buey}, {Carbillet}, {Carle}, {Cascone}, {Chapron}, {Charton},
  {Chauvin}, {Claudi}, {Costille}, {De Caprio}, {de Boer}, {Delboulb{\'e}},
  {Desidera}, {Dominik}, {Downing}, {Dupuis}, {Fabron}, {Fantinel}, {Farisato},
  {Feautrier}, {Fedrigo}, {Fusco}, {Gigan}, {Ginski}, {Girard}, {Giro},
  {Gisler}, {Gluck}, {Gry}, {Henning}, {Hubin}, {Hugot}, {Incorvaia}, {Jaquet},
  {Kasper}, {Lagadec}, {Lagrange}, {Le Coroller}, {Le Mignant}, {Le Ruyet},
  {Lessio}, {Lizon}, {Llored}, {Lundin}, {Madec}, {Magnard}, {Marteaud},
  {Martinez}, {Maurel}, {M{\'e}nard}, {Mesa}, {M{\"o}ller-Nilsson}, {Moulin},
  {Moutou}, {Orign{\'e}}, {Parisot}, {Pavlov}, {Perret}, {Pragt}, {Puget},
  {Rabou}, {Ramos}, {Reess}, {Rigal}, {Rochat}, {Roelfsema}, {Rousset}, {Roux},
  {Saisse}, {Salasnich}, {Santambrogio}, {Scuderi}, {Segransan}, {Sevin},
  {Siebenmorgen}, {Soenke}, {Stadler}, {Suarez}, {Tiph{\`e}ne}, {Turatto},
  {Udry}, {Vakili}, {Waters}, {Weber}, {Wildi}, {Zins}, \&
  {Zurlo}}]{Beuzit2019}
{Beuzit}, J.~L., {Vigan}, A., {Mouillet}, D., {et~al.} 2019, \aap, 631, A155

\bibitem[{{Bohn} {et~al.}(2020){Bohn}, {Southworth}, {Ginski}, {Kenworthy},
  {Maxted}, \& {Evans}}]{2020A&A...635A..73B}
{Bohn}, A.~J., {Southworth}, J., {Ginski}, C., {et~al.} 2020, \aap, 635, A73

\bibitem[{{Bradley} {et~al.}(2016){Bradley}, {Sipocz}, {Robitaille},
  {Tollerud}, {Deil}, {Vin{\'\i}cius}, {Barbary}, {G{\"u}nther}, {Bostroem},
  {Droettboom}, {Bray}, {Bratholm}, {Pickering}, {Craig}, {Pascual}, {Greco},
  {Donath}, {Kerzendorf}, {Littlefair}, {Barentsen}, {D'Eugenio}, \&
  {Weaver}}]{photutils}
{Bradley}, L., {Sipocz}, B., {Robitaille}, T., {et~al.} 2016, {Photutils:
  Photometry tools}

\bibitem[{{Cieza} {et~al.}(2009){Cieza}, {Padgett}, {Allen}, {McCabe},
  {Brooke}, {Carey}, {Chapman}, {Fukagawa}, {Huard}, {Noriga-Crespo},
  {Peterson}, \& {Rebull}}]{2009ApJ...696L..84C}
{Cieza}, L.~A., {Padgett}, D.~L., {Allen}, L.~E., {et~al.} 2009, \apjl, 696,
  L84

\bibitem[{{Dohlen} {et~al.}(2008){Dohlen}, {Langlois}, {Saisse}, {Hill},
  {Origne}, {Jacquet}, {Fabron}, {Blanc}, {Llored}, {Carle}, {Moutou}, {Vigan},
  {Boccaletti}, {Carbillet}, {Mouillet}, \& {Beuzit}}]{Dohlen2008}
{Dohlen}, K., {Langlois}, M., {Saisse}, M., {et~al.} 2008, in Society of
  Photo-Optical Instrumentation Engineers (SPIE) Conference Series, Vol. 7014,
  \procspie, 70143L

\bibitem[{{Dumusque} {et~al.}(2012){Dumusque}, {Pepe}, {Lovis},
  {S{\'e}gransan}, {Sahlmann}, {Benz}, {Bouchy}, {Mayor}, {Queloz}, {Santos},
  \& {Udry}}]{2012Natur.491..207D}
{Dumusque}, X., {Pepe}, F., {Lovis}, C., {et~al.} 2012, \nat, 491, 207

\bibitem[{{Duquennoy} \& {Mayor}(1991)}]{1991A&A...248..485D}
{Duquennoy}, A. \& {Mayor}, M. 1991, \aap, 500, 337

\bibitem[{{Eggenberger} {et~al.}(2007){Eggenberger}, {Udry}, {Chauvin},
  {Beuzit}, {Lagrange}, {S{\'e}gransan}, \& {Mayor}}]{2007A&A...474..273E}
{Eggenberger}, A., {Udry}, S., {Chauvin}, G., {et~al.} 2007, \aap, 474, 273

\bibitem[{{Eggenberger} {et~al.}(2011){Eggenberger}, {Udry}, {Chauvin},
  {Forveille}, {Beuzit}, {Lagrange}, \& {Mayor}}]{2011IAUS..276..409E}
{Eggenberger}, A., {Udry}, S., {Chauvin}, G., {et~al.} 2011, in IAU Symposium,
  Vol. 276, The Astrophysics of Planetary Systems: Formation, Structure, and
  Dynamical Evolution, ed. A.~{Sozzetti}, M.~G. {Lattanzi}, \& A.~P. {Boss},
  409--410

\bibitem[{{Esslinger} \& {Edmunds}(1998)}]{1998A&AS..129..617E}
{Esslinger}, O. \& {Edmunds}, M.~G. 1998, \aaps, 129, 617

\bibitem[{{Fontanive} {et~al.}(2019){Fontanive}, {Rice}, {Bonavita}, {Lopez},
  {Mu{\v{z}}i{\'c}}, {}, \& {Biller}}]{2019MNRAS.485.4967F}
{Fontanive}, C., {Rice}, K., {Bonavita}, M., {et~al.} 2019, \mnras, 485, 4967

\bibitem[{{Ford} {et~al.}(2000){Ford}, {Kozinsky}, \&
  {Rasio}}]{2000ApJ...535..385F}
{Ford}, E.~B., {Kozinsky}, B., \& {Rasio}, F.~A. 2000, \apj, 535, 385

\bibitem[{{Foreman-Mackey} {et~al.}(2013){Foreman-Mackey}, {Hogg}, {Lang}, \&
  {Goodman}}]{2013PASP..125..306F}
{Foreman-Mackey}, D., {Hogg}, D.~W., {Lang}, D., \& {Goodman}, J. 2013, \pasp,
  125, 306

\bibitem[{{Gaia Collaboration}(2018)}]{2018yCat.1345....0G}
{Gaia Collaboration}. 2018, VizieR Online Data Catalog, I/345

\bibitem[{{Girardi} {et~al.}(2000){Girardi}, {Bressan}, {Bertelli}, \&
  {Chiosi}}]{Girardi2000}
{Girardi}, L., {Bressan}, A., {Bertelli}, G., \& {Chiosi}, C. 2000, \aaps, 141,
  371

\bibitem[{{Guyon}(2018)}]{2018ARA&A..56..315G}
{Guyon}, O. 2018, \araa, 56, 315

\bibitem[{{Haisch} {et~al.}(2001){Haisch}, {Lada}, \&
  {Lada}}]{2001ApJ...553L.153H}
{Haisch}, Karl~E., J., {Lada}, E.~A., \& {Lada}, C.~J. 2001, \apjl, 553, L153

\bibitem[{{Harakawa} {et~al.}(2015){Harakawa}, {Sato}, {Omiya}, {Fischer},
  {Hori}, {Ida}, {Kambe}, {Yoshida}, {Izumiura}, {Koyano}, {Nagayama},
  {Shimizu}, {Okada}, {Okita}, {Sakamoto}, \& {Yamamuro}}]{2015ApJ...806....5H}
{Harakawa}, H., {Sato}, B., {Omiya}, M., {et~al.} 2015, \apj, 806, 5

\bibitem[{{Holberg} \& {Bergeron}(2006)}]{holberg2006}
{Holberg}, J.~B. \& {Bergeron}, P. 2006, AJ, 132, 1221

\bibitem[{{Houk}(1982)}]{Houk1982}
{Houk}, N. 1982, {Michigan Catalogue of Two-dimensional Spectral Types for the
  HD stars. Volume\_3. Declinations -40.0 to -26.0.}

\bibitem[{{Houk} \& {Smith-Moore}(1988)}]{1988mcts.book.....H}
{Houk}, N. \& {Smith-Moore}, M. 1988, {Michigan Catalogue of Two-dimensional
  Spectral Types for the HD Stars. Volume 4, Declinations -26$^\circ$.0 to
  -12$^\circ$.0.}, Vol.~4

\bibitem[{{Hunter}(2007)}]{Matplotlib}
{Hunter}, J.~D. 2007, Computing in Science and Engineering, 9, 90

\bibitem[{{Jones} {et~al.}(2015){Jones}, {Jenkins}, {Rojo}, {Olivares}, \&
  {Melo}}]{Jones2015}
{Jones}, M.~I., {Jenkins}, J.~S., {Rojo}, P., {Olivares}, F., \& {Melo},
  C.~H.~F. 2015, \aap, 580, A14

\bibitem[{{Kowalski} \& {Saumon}(2006)}]{kowalski2006}
{Kowalski}, P.~M. \& {Saumon}, D. 2006, ApJL, 651, L137

\bibitem[{{Kozai}(1962)}]{1962AJ.....67..591K}
{Kozai}, Y. 1962, \aj, 67, 591

\bibitem[{{Kraus} {et~al.}(2012){Kraus}, {Ireland}, {Hillenbrand}, \&
  {Martinache}}]{2012ApJ...745...19K}
{Kraus}, A.~L., {Ireland}, M.~J., {Hillenbrand}, L.~A., \& {Martinache}, F.
  2012, \apj, 745, 19

\bibitem[{{Lo Curto} {et~al.}(2013){Lo Curto}, {Mayor}, {Benz}, {Bouchy},
  {H{\'e}brard}, {Lovis}, {Moutou}, {Naef}, {Pepe}, {Queloz}, {Santos},
  {Segransan}, \& {Udry}}]{LoCurto2013}
{Lo Curto}, G., {Mayor}, M., {Benz}, W., {et~al.} 2013, \aap, 551, A59

\bibitem[{{Maire} {et~al.}(2016){Maire}, {Langlois}, {Dohlen}, {Lagrange},
  {Gratton}, {Chauvin}, {Desidera}, {Girard}, {Milli}, {Vigan}, {Zins},
  {Delorme}, {Beuzit}, {Claudi}, {Feldt}, {Mouillet}, {Puget}, {Turatto}, \&
  {Wildi}}]{Maire2016}
{Maire}, A.-L., {Langlois}, M., {Dohlen}, K., {et~al.} 2016, in Society of
  Photo-Optical Instrumentation Engineers (SPIE) Conference Series, Vol. 9908,
  \procspie, 990834

\bibitem[{{Mawet} {et~al.}(2014){Mawet}, {Milli}, {Wahhaj}, {Pelat}, {Absil},
  {Delacroix}, {Boccaletti}, {Kasper}, {Kenworthy}, {Marois}, {Mennesson}, \&
  {Pueyo}}]{2014ApJ...792...97M}
{Mawet}, D., {Milli}, J., {Wahhaj}, Z., {et~al.} 2014, \apj, 792, 97

\bibitem[{{Mel{\'e}ndez} {et~al.}(2017){Mel{\'e}ndez}, {Bedell}, {Bean},
  {Ram{\'\i}rez}, {Asplund}, {Dreizler}, {Yan}, {Shi}, {Lind}, {Ferraz-Mello},
  {Galarza}, {dos Santos}, {Spina}, {Maia}, {Alves-Brito}, {Monroe}, \&
  {Casagrande}}]{Melendez2017}
{Mel{\'e}ndez}, J., {Bedell}, M., {Bean}, J.~L., {et~al.} 2017, \aap, 597, A34

\bibitem[{{Mugrauer}(2019)}]{Mugrauer2019}
{Mugrauer}, M. 2019, \mnras, 490, 5088

\bibitem[{{Mugrauer} \& {Ginski}(2015)}]{2015MNRAS.450.3127M}
{Mugrauer}, M. \& {Ginski}, C. 2015, \mnras, 450, 3127

\bibitem[{{Nelson}(2000)}]{2000ApJ...537L..65N}
{Nelson}, A.~F. 2000, \apjl, 537, L65

\bibitem[{{Ngo} {et~al.}(2017){Ngo}, {Knutson}, {Bryan}, {Blunt}, {Nielsen},
  {Batygin}, {Bowler}, {Crepp}, {Hinkley}, {Howard}, \& {Mawet}}]{Ngo2017}
{Ngo}, H., {Knutson}, H.~A., {Bryan}, M.~L., {et~al.} 2017, \aj, 153, 242

\bibitem[{{Ngo} {et~al.}(2016){Ngo}, {Knutson}, {Hinkley}, {Bryan}, {Crepp},
  {Batygin}, {Crossfield}, {Hansen}, {Howard}, {Johnson}, {Mawet}, {Morton},
  {Muirhead}, \& {Wang}}]{2016ApJ...827....8N}
{Ngo}, H., {Knutson}, H.~A., {Hinkley}, S., {et~al.} 2016, \apj, 827, 8

\bibitem[{{Ngo} {et~al.}(2015){Ngo}, {Knutson}, {Hinkley}, {Crepp}, {Bechter},
  {Batygin}, {Howard}, {Johnson}, {Morton}, \&
  {Muirhead}}]{2015ApJ...800..138N}
{Ngo}, H., {Knutson}, H.~A., {Hinkley}, S., {et~al.} 2015, \apj, 800, 138

\bibitem[{Oliphant(2006)}]{oliphant2006guide}
Oliphant, T.~E. 2006, A guide to NumPy, Vol.~1 (Trelgol Publishing USA)

\bibitem[{{Pascucci} {et~al.}(2008){Pascucci}, {Apai}, {Hardegree-Ullman},
  {Kim}, {Meyer}, \& {Bouwman}}]{2008ApJ...673..477P}
{Pascucci}, I., {Apai}, D., {Hardegree-Ullman}, E.~E., {et~al.} 2008, \apj,
  673, 477

\bibitem[{{Pecaut} \& {Mamajek}(2013)}]{pecaut2013}
{Pecaut}, M.~J. \& {Mamajek}, E.~E. 2013, ApJS, 208, 9

\bibitem[{{Raghavan} {et~al.}(2010){Raghavan}, {McAlister}, {Henry}, {Latham},
  {Marcy}, {Mason}, {Gies}, {White}, \& {ten Brummelaar}}]{2010ApJS..190....1R}
{Raghavan}, D., {McAlister}, H.~A., {Henry}, T.~J., {et~al.} 2010, \apjs, 190,
  1

\bibitem[{{Roell} {et~al.}(2012){Roell}, {Neuh{\"a}user}, {Seifahrt}, \&
  {Mugrauer}}]{2012A&A...542A..92R}
{Roell}, T., {Neuh{\"a}user}, R., {Seifahrt}, A., \& {Mugrauer}, M. 2012, \aap,
  542, A92

\bibitem[{{Rosotti} \& {Clarke}(2018)}]{2018MNRAS.473.5630R}
{Rosotti}, G.~P. \& {Clarke}, C.~J. 2018, \mnras, 473, 5630

\bibitem[{{Skrutskie} {et~al.}(2006){Skrutskie}, {Cutri}, {Stiening},
  {Weinberg}, {Schneider}, {Carpenter}, {Beichman}, {Capps}, {Chester},
  {Elias}, {Huchra}, {Liebert}, {Lonsdale}, {Monet}, {Price}, {Seitzer},
  {Jarrett}, {Kirkpatrick}, {Gizis}, {Howard}, {Evans}, {Fowler}, {Fullmer},
  {Hurt}, {Light}, {Kopan}, {Marsh}, {McCallon}, {Tam}, {Van Dyk}, \&
  {Wheelock}}]{2006AJ....131.1163S}
{Skrutskie}, M.~F., {Cutri}, R.~M., {Stiening}, R., {et~al.} 2006, \aj, 131,
  1163

\bibitem[{{Takeda} \& {Rasio}(2005)}]{2005ApJ...627.1001T}
{Takeda}, G. \& {Rasio}, F.~A. 2005, \apj, 627, 1001

\bibitem[{{Thebault} \& {Haghighipour}(2015)}]{2015pes..book..309T}
{Thebault}, P. \& {Haghighipour}, N. 2015, {Planet Formation in Binaries},
  309--340

\bibitem[{{Tremblay} {et~al.}(2011){Tremblay}, {Bergeron}, \&
  {Gianninas}}]{tremblay2011}
{Tremblay}, P.-E., {Bergeron}, P., \& {Gianninas}, A. 2011, ApJ, 730, 128

\bibitem[{{van Holstein} {et~al.}(2020){van Holstein}, {Girard}, {de Boer},
  {Snik}, {Milli}, {Stam}, {Ginski}, {Mouillet}, {Wahhaj}, {Schmid}, {Keller},
  {Langlois}, {Dohlen}, {Vigan}, {Pohl}, {Carbillet}, {Fantinel}, {Maurel},
  {Orign{\'e}}, {Petit}, {Ramos}, {Rigal}, {Sevin}, {Boccaletti}, {Le
  Coroller}, {Dominik}, {Henning}, {Lagadec}, {M{\'e}nard}, {Turatto}, {Udry},
  {Chauvin}, {Feldt}, \& {Beuzit}}]{vanHolstein2020}
{van Holstein}, R.~G., {Girard}, J.~H., {de Boer}, J., {et~al.} 2020, \aap,
  633, A64

\bibitem[{{van Leeuwen}(2007)}]{Leeuwen2007}
{van Leeuwen}, F. 2007, \aap, 474, 653

\bibitem[{{Virtanen} {et~al.}(2020){Virtanen}, {Gommers}, {Oliphant},
  {Haberland}, {Reddy}, {Cournapeau}, {Burovski}, {Peterson}, {Weckesser},
  {Bright}, {van der Walt}, {Brett}, {Wilson}, {Jarrod Millman}, {Mayorov},
  {Nelson}, {Jones}, {Kern}, {Larson}, {Carey}, {Polat}, {Feng}, {Moore}, {Vand
  erPlas}, {Laxalde}, {Perktold}, {Cimrman}, {Henriksen}, {Quintero}, {Harris},
  {Archibald}, {Ribeiro}, {Pedregosa}, {van Mulbregt}, \&
  {Contributors}}]{2020SciPy-NMeth}
{Virtanen}, P., {Gommers}, R., {Oliphant}, T.~E., {et~al.} 2020, Nature
  Methods, 17, 261

\bibitem[{{Wang} {et~al.}(2015){Wang}, {Fischer}, {Xie}, \&
  {Ciardi}}]{2015ApJ...813..130W}
{Wang}, J., {Fischer}, D.~A., {Xie}, J.-W., \& {Ciardi}, D.~R. 2015, \apj, 813,
  130

\bibitem[{{Wenger} {et~al.}(2000){Wenger}, {Ochsenbein}, {Egret}, {Dubois},
  {Bonnarel}, {Borde}, {Genova}, {Jasniewicz}, {Lalo{\"e}}, {Lesteven}, \&
  {Monier}}]{Wenger2000}
{Wenger}, M., {Ochsenbein}, F., {Egret}, D., {et~al.} 2000, \aaps, 143, 9

\end{thebibliography}

\end{document}